\documentclass[longauth]{aa} 
\usepackage{graphicx}
\usepackage{txfonts}
\begin{document}
   \title{The VVV Templates Project \\Towards an automated classification of VVV light-curves}
   \subtitle{I. Building a database of stellar variability in the near-infrared}

   \author{R. Angeloni\inst{1,2,3}\thanks{rangelon@astro.puc.cl},
	  R. Contreras Ramos\inst{1,4},
	  M. Catelan\inst{1,2,4},
	  I. D\'ek\'any\inst{4,1},
	  F. Gran\inst{1,4},
	  J. Alonso-Garc\'ia\inst{1,4},
	  M. Hempel\inst{1,4},
	  C. Navarrete\inst{1,4},
	  H. Andrews\inst{1,6},
	  A. Aparicio\inst{7,21},
	  J.~C.~Beam\'in\inst{1,4,8},
          C. Berger\inst{5},
          J. Borissova\inst{9,4},
          C. Contreras Pe\~{n}a\inst{10},
          A. Cunial\inst{11,12},
          R. de Grijs\inst{13,14},
          N. Espinoza\inst{1,15,4},
          S. Eyheramendy\inst{15,4},
          C. E. Ferreira Lopes\inst{16},
          M. Fiaschi\inst{12},
          G. Hajdu\inst{1,4},
          J. Han\inst{17},
          K. G. He{\l}miniak\inst{18,19},
          A. Hempel\inst{20},
          S. L. Hidalgo\inst{7,21},
          Y. Ita\inst{22},
          Y. -B. Jeon\inst{23},
          A. Jord\'an\inst{1,2,4},
          J. Kwon\inst{24},
          J. T. Lee\inst{17},
          E. L. Mart\'{\i}n\inst{25},
          N. Masetti\inst{26},
          N. Matsunaga\inst{27},
          A. P. Milone\inst{28},
          D. Minniti\inst{1,4},
          L. Morelli\inst{11,12},
          F. Murgas\inst{7,21},
          T. Nagayama\inst{29},
          C. Navarro\inst{9,4},
          P. Ochner\inst{12},
          P. P\'{e}rez\inst{30},
          K. Pichara\inst{5,4},
          A. Rojas-Arriagada\inst{31},
          J. Roquette\inst{32},
          R. K. Saito\inst{33},
          A. Siviero\inst{12},
          J. Sohn\inst{17},
          H. -I. Sung\inst{23},
          M. Tamura\inst{27,24},
          R. Tata\inst{7},
          L. Tomasella\inst{12},
          B. Townsend\inst{1,4},
        \and
          P. Whitelock\inst{34,35}
         }

   \institute{Instituto de Astrof\'isica, Pontificia Universidad Cat\'olica de Chile, Av. Vicu\~{n}a Mackenna 4860, 782-0436 Macul, Santiago, Chile
         \and 
             Centro de Astro-Ingenier\'{\i}a, Pontificia Universidad Cat\'olica de Chile, Santiago, Chile
         \and
             Max-Planck-Institut f\"{u}r Astronomie, K\"{o}nigstuhl 17, 69117 Heidelberg, Germany
         \and
            Millennium Institute of Astrophysics, Santiago, Chile
         \and
            Computer Science Department, Pontificia Universidad Cat\'olica de Chile, Santiago, Chile 
         \and
            Leiden Observatory, Leiden, The Netherlands
         \and 
            Instituto de Astrof\'{\i}sica de Canarias (IAC), V\'{\i}a L\'actea s/n, E-38200 La Laguna, Tenerife, Canary Islands, Spain
         \and
            European Southern Observatory, Av. Alonso de C\'{o}rdoba 3107, Casilla 19001, Santiago, Chile 
         \and    
             Instituto de F\'{\i}sica y Astronom\'{\i}a, Universidad de Valpara\'{\i}so, Av. Gran Breta\~{n}a 1111, Playa Ancha, Casilla 5030, Chile
          \and
             Centre for Astrophysics Research, University of Hertfordshire, Hatfield, AL10 9AB, UK   
         \and
             Dipartimento di Fisica e Astronomia "Galileo Galilei", Universit\`{a} di Padova, Vicolo dell'Osservatorio 3, Padova I-35122, Italy
          \and  
             INAF -- Osservatorio Astronomico di Padova, vicolo dell'Osservatorio 5, 35122 Padova, Italy  
          \and  
             Kavli Institute for Astronomy and Astrophysics, Peking University, Yi He Yuan Lu 5, Hai Dian District, Beijing 100871, China
          \and 
             Department of Astronomy, Peking University, Yi He Yuan Lu 5, Hai Dian District, Beijing 100871, China
          \and   
             Facultad de Matematicas, Pontificia Universidad Cat\'olica de Chile, Santiago, Chile
          \and  
            Universidade Federal do Rio Grande do Norte, Campus Universit\'{a}rio Lagoa Nova, CEP 59078-970, Natal, Brasil 
         \and
             Korea National University of Education, San 7, GangNaeMyeon, CheongWonGun, ChungBuk 363-791, The Republic of Korea 
         \and 
             Subaru Telescope, National Astronomical Observatory of Japan, 650 North Aohoku Place, Hilo, HI 96720, USA
         \and
            Nicolaus Copernicus Astronomical Center, Department of Astrophysics, ul. Rabia\'{n}ska 8, 87-100 Toru\'{n}, Poland 
         \and  
             Departamento de Ciencias F\'{\i}sicas, Universidad Andr\'{e}s Bello, Av. Rep\'{u}blica 252, Santiago, Chile
         \and
            Department of Astrophysics, University of La Laguna, V\'{\i}ía L\'{a}ctea s/n, E-38200 La Laguna, Tenerife, Canary Islands, Spain 
         \and 
             Astronomical Institute, Graduate School of Science, Tohoku University, 6-3 Aramaki Aoba, Aoba-ku, Sendai, Miyagi 980-8578, Japan
         \and
	     Korea Astronomy and Space Science Institute, Daedeokdae-ro, Yuseong-gu, Daejeon, Republic of Korea 305-348
         \and  
              National Astronomical Observatory, Osawa 2-21-1, Mitaka, Tokyo, Japan
         \and
             Centro de Astrobiolog\'{\i}a (CSIC-INTA), Ctra. Ajalvir km 4, 28850 Torrej\'on de Ardoz, Madrid, Spain
         \and
            INAF -- Istituto di Astrofisica Spaziale e Fisica Cosmica di Bologna, via Gobetti 101, I-40129 Bologna, Italy  
         \and
            Department of Astronomy, Graduate School of Science, The University of Tokyo, 7-3-1 Hongo, Bunkyo-ku, Tokyo 113-0033, Japan
        \and   
             Research School of Astronomy \& Astrophysics, Australian National University, Mt. Stromlo Observatory, ACT 2611, Australia
        \and  
             Department of Astrophysics, Nagoya University, Furo-cho, Chikusa-ku, Nagoya 464-8602, Japan
        \and
            Departamento de Ciencias de la Computaci\'{o}n, Universidad de Chile, Casilla 2777, Av. Blanco Encalada 2120, Santiago, Chile
        \and
        Laboratoire Lagrange (UMR 7293), Universit\'e  Nice Sophia Antipolis, CNRS, Observatoire de la C\^ote d'Azur, BP 4229, 06304 Nice, France
        \and
        Departamento de F\'{\i}sica - ICEx - UFMG, Av. Ant\^{o}nio Carlos, 6627, 30270-901, Belo Horizonte, MG, Brazil
        \and
            Universidade Federal de Sergipe, Departamento de F\'{\i}sica, Av. Marechal Rondon s/n, 49100-000 S\~{a}o Crist\'{o}v\~{a}o, SE, Brazil
        \and
            South African Astronomical Observatory, PO Box 9, 7935 Observatory, South Africa
        \and 
            Astrophysics, Cosmology and Gravity Centre, Astronomy Department, University of Cape Town, Rondebosch 7701, South Africa
             }

   \date{2013;2014}

\authorrunning{Angeloni et al.}
\titlerunning{The VVV Templates Project}
   
 
  \abstract
   {The Vista Variables in the V\'ia L\'actea (VVV) ESO Public Survey is a variability survey of the Milky Way bulge and an adjacent section of the disk carried out from 2010 on ESO Visible and Infrared Survey Telescope for Astronomy (VISTA). The VVV survey will eventually deliver a deep near-IR atlas with photometry and positions in five passbands ($ZYJHK_S$) and a catalogue of 1-10 million variable point sources -- mostly unknown -- that require classifications.}
   {The main goal of the VVV Templates Project, which we introduce in this work, is to develop and test the machine-learning algorithms for the automated classification of the VVV light-curves. As VVV is the first massive, multi-epoch survey of stellar variability in the near-infrared, the template light-curves that are required for training the classification algorithms are not available. In the first paper of the series we  describe the construction of this comprehensive database of infrared stellar variability.}
   {First, we performed a systematic search in the literature and public data archives; second, we coordinated a worldwide observational campaign; and third, we exploited the VVV variability database itself on (optically) well-known stars to gather high-quality infrared light-curves of several hundreds of variable stars.}
   {We have now collected a significant (and still increasing) number of infrared template light-curves. This database will be used as a training-set for the machine-learning algorithms that will automatically classify the light-curves produced by VVV. The results of such an automated classification will be covered in forthcoming papers of the series.}
  {}


   \maketitle
%

\section{Introduction}\label{sect:intro}
The \textit{Vista Variables in the V\'ia L\'actea} (VVV) ESO Public Survey is an infrared (IR) variability survey of the Milky Way bulge and an adjacent section of the mid-plane (Minniti et al. 2010; Saito et al. 2010, 2012; Catelan et al. 2011, 2013) carried out on the 4m Visible and Infrared Survey Telescope for Astronomy (VISTA) at ESO. The permanently mounted instrumentation, VIRCAM (Dalton et al. 2006), comprises a wide-field near-IR mosaic of 16 2048$\times$2048 detectors. It is sensitive over the spectral region 0.8-2.5 $\mu$m, and with an average scale of $0.34"$/pixel and the total (effective) field of view (FOV) is 1.1$\times$1.5 square degrees. The survey observations started in 2010 and are expected to end in late 2016 or early 2017. The survey covers a total area of 562 square degrees, monitoring a billion point sources. The final products will be a deep near-IR (0.9-2.5 $\mu$m) atlas with photometry and positions in five passbands ($ZYJHK_S$) and a catalogue of a few million variable 
point sources that require classifying in an 
automated way. The VVV Survey is proving 
to be every bit as much of a gold-mine as anticipated, having already produced a number of scientific results on the structure of the Milky Way bulge (D\'ek\'any et al. 2013) and disk (He{\l}miniak et al. 2013; Soto et al. 2013), extinction maps (Gonzalez et al. 2013), star clusters (Hempel 2013; Borissova et al. 2011; Minniti et al. 2011; Moni-Bidin et al. 2011), high proper motions stars (Ivanov et al. 2013), brown dwarfs (Beam\'in et al. 2013), planetary nebulae (Weidmann et al. 2013), novae (Saito et al. 2013), as well as variable stars (Catelan et al. 2013). 

What makes VVV unique with respect to other (both past and ongoing) very large variability surveys is that it is performed in the near-IR. Although observations in the near-IR have fundamental advantages, mostly due to the ability to probe deeper into the heavily reddened regions of the Galactic bulge and plane, observations in this spectral region also present significant challenges. In particular, the high-quality light-curves that are needed as templates for training and testing the automated variable star classification algorithms are not available, yet.
This is very different from the situation in the optical, for which numerous light-curves of template quality have been published (see e.g. Debosscher et al. 2007; Blomme et al. 2011; Dubath et al. 2011). The currently available near-IR templates are inadequate for statistical purposes, since they do not properly sample the large variety of light-curve shapes encountered for many of the most important variability classes (e.g. Eclipsing Binaries; see also Table \ref{tab:listref}).  Furthermore, several classes of variables have not been observed in a sufficiently extensive way in the near-IR, so their IR behaviour is still largely unknown (e.g. SX Phe stars). 

In order to properly classify as a template, a light-curve should meet some minimum observational criteria. The most important features are (1) the observations must be very precise, with the relative error in the photometry less than about 1/10 of the light-curve amplitude and  (2) the light-curve itself must be complete, without any significant gaps in the phase coverage. This is of great importance to properly compute accurate Fourier decomposition parameters for the light-curves, as required for the automated classification algorithms. Nonetheless, sometimes compromises must be made and constraints reduced on specific template-quality definitions given the realities of near-IR observations when dealing with the astrophysical (i.e. smaller pulsational amplitudes) and technical (e.g. strong and highly variable sky background) challenges.

Current estimates based on the analysis of the increasingly larger VVV dataset suggest that the final number of truly variable stars observed by VVV will be between 10$^6$ and 10$^7$ (Catelan et al. 2013). When dealing with such large numbers of objects, automated methods in the data selection, analysis, and classification have obviously to be searched for, tested, and optimized.
In recent years there has been an increasing interest in the application of artificial intelligence algorithms to astronomical research, and machine learning techniques have been applied to variable star classification. For instance, Debosscher et al. (2007) explored several classification techniques, quantitatively comparing the performance (e.g. in terms of computational time) and final results (e.g. in terms of accuracy) of different classifiers with their corresponding learning algorithms. More recently, other studies have focused on specific methodologies, with the implicit goal of finding the best compromise between robustness and speed of different machine-learning algorithms (e.g. Blomme et al. 2011; Dubath et al. 2011; Richards et al. 2011; Pichara et al. 2012; Pichara \& Protopapas 2013). 

Whichever specific algorithm is chosen, the general idea behind these supervised machine learning methodologies is to create a function, the classifier, able to infer the most probable label for an object (in our case, the variability class to which a star belongs) on the basis of what is learned by the analysis of inputs (light-curve features) from a training set (a collection of high-quality light-curves of previously classified variable stars).
In the most general framework, the sequence of steps to be performed can be considered 1) build a training set; 2) determine the input feature representation of the learned function;  3) determine the nature of the classifier with its corresponding learning algorithm (e.g. artificial neural network, support vector machines, tree-based methods, etc.); 4) run the classifier on the gathered test set; and 5) evaluate the accuracy of the learned function.

In astronomical terms, this would translate into: 1) collecting high-quality template light-curves for a significant sample of stars taken to be representative of the different variability classes under study; 2) identifying and extracting the most informative features that best describe the temporal series (e.g. periods and harmonics as derived from Fourier analysis, amplitudes, colours, skewness, etc.) from the training data gathered at step 1; 3) selecting the supervised machine-learning technique thought to perform most effectively for the particular science case; 4) feeding the algorithms chosen at step 3 with the light-curve features extracted at step 2, using the information on well-known variables from the training-set to search for and label unknown variable stars in the test-set, i.e. in the data archive; and 5) evaluating the fraction of correctly classified variable stars (e.g. through a so-called confusion matrix; Debosscher et al. 2007).\\

Because VVV is the first massive, multi-epoch survey dedicated to stellar variability in the near-IR, the first problem we had to face was the construction of a proper training set. This is a database of high-quality (i.e. template-like) near-IR light-curves for a significant sample of stars representative of the various variability classes to be observed by VVV. We have named this task the VVV Templates Project\footnote{\tt http://www.vvvtemplates.org/}.

In the present paper, the first of a series, we introduce the VVV Templates Project by describing the first step of the aforementioned working strategy, namely, the coordinated program of building the first-ever database of high-quality light-curves in the near-IR. First, we describe in Sect. \ref{sect:antes} the data mining of high-quality time-series data in both the literature and in public IR archives. Then, in Sect. \ref{sect:network}, we outline our own observing project whose aim was essentially to increase both the number and the types of variables characterized in the near-IR. In addition, VVV itself is obtaining data within the Galactic bulge and disk where accurate variable star classifications have already been made by optical surveys (e.g. 
 MACHO -- Alcock et al. 1992 and OGLE -- Udalski 2003): this is the topic of Sect. \ref{sect:vvv}. Eventually, concluding remarks are given in Sect. \ref{sec:cr}.

\section{Near-IR data from the literature and public archives}\label{sect:antes}

\subsection{Digging into the literature}
We extensively explored the available literature looking for papers that presented high-quality IR time-series data, that we digitized following three main methods.
1)~When data were reported in table format but without an accompanying digital version, Optical Character Recognition (OCR) was used.
2)~When the data points were available only as plots or figures, the actual data points were extracted using the DataThief (Tummers 2006)
program. 3)~Finally, for the more recent references, the ASCII tables were directly extracted from the online versions of the articles.

These digitized light-curves belong to multiple groups like Classical Cepheids, Eclipsing Binaries, Miras, RR Lyrae, R Coronae Borealis stars, and others. Table \ref{tab:listref} lists the papers from which we could extract template-quality IR time-series data. 

\begin{table*}
\caption{Stellar variability classes with high-quality IR time-series data from the literature, the GCVS designation (in brackets), the number of retrieved light-curves (NLC), and the corresponding reference source(s).\label{tab:listref}}
\centering
    \begin{tabular}{ccccccccc}
	\hline \hline
         Variability class   & NLC    & References \\
 \hline
        Cataclysmic variables (CV)  & 4 & Szkody et al. (1983); Bailey et al. (1985); Ribeiro \& Baptista (2011)\\
        Classical Cepheids (DCEP) & $\sim$240 & P04; G13; MP11; R12(a,b); M14$^a$\\	
        Anomalous Cepheids (BLBOO) & 48 & Ripepi et al. (2014)\\
        Eclipsing Binaries (EA+EB+EW) & 13 & A90; A95: A02; B81; L97; L02; L04; L09; MP08; S82; S84; C04$^b$\\ 
        Long-Period Variables (LPV)  & $\sim$260 & Olivier et al. (2001); Smith et al. (2002); Whitelock et al. (2000, 2006)\\ 
        Quasi-Stellar Objects (QSO) & $\sim$70 & Cioni et al. (2013)\\
        R Coronae Borealis (RCB)   & 12 & Feast et al. (1997)\\
        RR Lyrae, type RRab (RRab)  & 44 & B92; C92; DP05; F89; F90a; J88; J92; J96; S89; S93; ST92; R12(a,b); M14; Sz14$^c$\\
        RR Lyrae, type RRc (RRc) & 8 & F90b; D05; J96; R12(a,b); M14$^d$\\
        T Tauri stars (INT)  & 1 & Lorenzetti et al. (2007)\\
        Wolf-Rayet stars (WR) & 6 & Van der Hucht et al. (2001); Williams et al. (1990, 2012, 2013) \\
        \hline
    \end{tabular}
     \begin{flushleft}
     {$a$}: P04: Persson et al. (2004); G13: Groenewegen (2013); MP11: Monson \& Pierce (2011); Ripepi et al. (2012a,b); Moretti et al. (2014).\\
     {$b$}: A90: Arevalo \& L\'azaro (1990); A95: Arevalo et al. (1995); A02: Arevalo et al. (2002); B81: Bailey et al. (1981); L97: L\'azaro et al. (1997); L02: L\'azaro et al. (2002); L04: L\'azaro et al. (2004); L09: L\'azaro et al. (2009); MP08: Mart\'inez-Pais et al. (2008); S82: Sherrington et al. (1982); S84: Sherrington et al. (1984); C04: Covino et al. (2004).\\
     {$c$}: B92: Barnes et al. (1992); C92: Cacciari et al. (1992); DP05: Del Principe et al. (2005); F89: Fernley et al. (1989); F90: Fernley et al. (1990a); J88: Jones et al. (1988); J92: Jones et al. (1992); J96: Jones et al. (1996); S89: Skillen et al. (1989); S93: Skillen et al. (1993); ST92: Storm et al. (1992); R12: Ripepi et al. (2012); M14: Moretti et al. (2014); Sz14: Szab{\'o} et al. (2014).\\
     {$d$}: F90: Fernley et al. (1990b); D05: Del Principe et al. (2005); J96: Jones et al. (1996); R12(a,b): Ripepi et al. (2012a,b); M14: Moretti et al. (2014).

    \end{flushleft}
\end{table*}


\subsection{Digging into public databases}
Public astronomical archives represent another source of high-quality data. Time-series data are generated as by-products of various observational programs, and an increasing fraction of these can be effectively exploited via the Virtual Observatory.

\subsubsection{2MASS}
The Two Micron All Sky Survey (2MASS, Skrutskie et al. 2006) was a near-IR ($JHK_S$)
imaging survey covering most of the sky. The final products of the survey are the
point-source and extended-source catalogues containing unique measurements of detected
astronomical objects. During the survey 35 standard fields were regularly monitored
each night for calibration purposes. The individual measurements of objects detected
for these exposures are available on the NASA/IPAC Infrared Science Archive
\footnote{\tt http://irsa.ipac.caltech.edu/} among the Ancillary Data Products of the
2MASS survey. The 2MASS observations and data products are described in Cutri et al. (2003).

We downloaded all photometric data of the objects detected in the calibration
fields from the 2MASS Calibration Point Source Working Database through IPAC. The
number of detections for each of the individual objects in the calibration fields
ranges from several hundreds to $\sim$3000, depending on field, position, and magnitude (see also Quillen et al. 2014).

\begin{figure}
   \centering
   \includegraphics[width=0.48\textwidth]{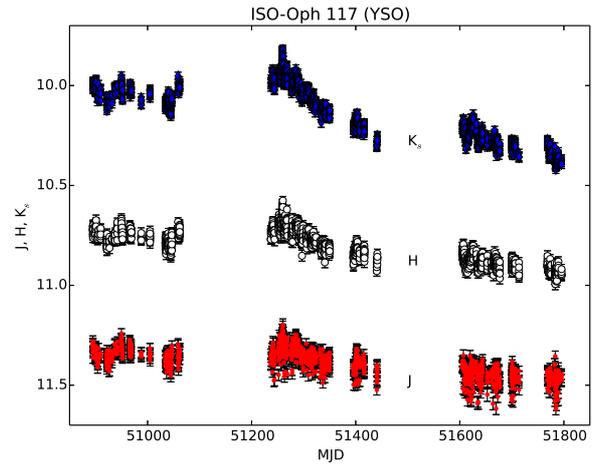}
   \caption{$JHK_S$ light-curve of a young stellar object (YSO) recovered from the 2MASS calibration fields.}
         \label{fig:2massgergely}
\end{figure}

The detections were cross-correlated with the General Catalogue of Variable Stars (GCVS; Samus et al. 2009a), the ASAS Catalogue of Variable Stars (ACVS; Pojmansky 2002), and the International Variable Star Index (VSX; Watson et al. 2014). The total number of matches, taking into account records in common among the catalogues, is 187. Each light-curve was inspected individually to determine its quality; 89 of them show light-curve variations and have the quality required to serve as templates (see Fig. \ref{fig:2massgergely} for an example of a $JHK_S$ light-curve of a young stellar object).

In future, we plan to cross-correlate 2MASS data with other variability archives, and also to do a general variable search using the 2MASS light-curves, similarly to what we are currently doing with the WFCAM data (Sect. \ref{sect:wfcam}).

\subsubsection{The SIRIUS/IRSF Variability Surveys}\label{sect:matsunaga}
The Infrared Survey Facility (IRSF; Glass \& Nagata 2000) is a joint Japanese/South African project\footnote{\tt http://www.saao.ac.za/$\sim$isg/irsf.html} that has been operational since 2000 at the South African Astronomical Observatory (SAAO). It consists of a 1.4m telescope and a simultaneous three-channel $JHK_S$ imager called SIRIUS (see also Sect. \ref{sect:irsf}). Constructed at Nagoya University, SIRIUS uses three 1024$\times$1024 HgCdTe arrays mapping an overall FOV of approximately 7.7$\times$7.7 arcminutes$^2$ with a pixel-scale of 0.45 arcseconds/pixel (Nagayama et al. 2003; see also Kato et al. 2007 for the data reduction pipeline).

\begin{figure}
   \centering
   \includegraphics[width=0.48\textwidth]{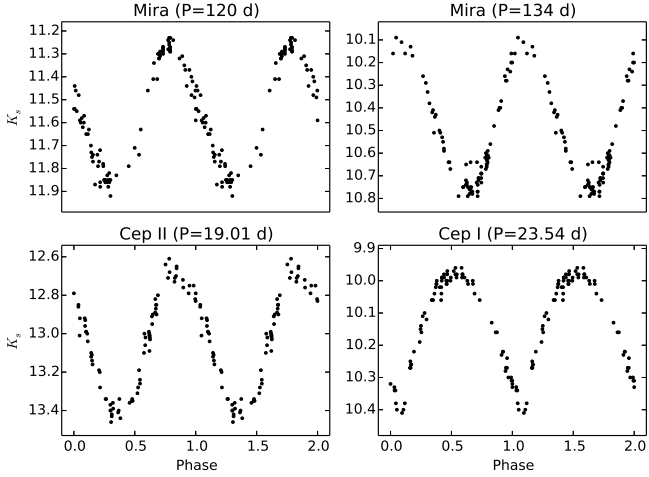}
   \caption{$K_S$-band light-curves of two Mira (top panels) and two (a Classical and a Type II, bottom panels) Cepheid variables obtained in the course of the SIRIUS/IRSF Variability Surveys.}
         \label{fig:matsunaga}
\end{figure}

There have been several extensive IRSF/SIRIUS surveys of variable stars in various regions:
the Magellanic Clouds (Ita et al. 2004a,b, 2009), the Galactic Center (Matsunaga et al. 2009, 2011, 2013), globular clusters (Matsunaga et al. 2006, Sloan et al. 2010), and Local Group dwarf galaxies (Whitelock et al. 2013 and references therein).
Considering the numbers of monitoring epochs, the surveys discussed below provide us with particularly useful datasets to construct template light-curves. Matsunaga et al. (2006) obtained well-sampled light-curves for dozens of Type II Cepheids in various globular clusters,
and the data are available online. Matsunaga et al. (2009, 2013) published a large set of light-curves including more than 1000 long-period variables, 3 classical Cepheids, and 16 Type II Cepheids based on their 8 year survey towards 20 arcmin times 30 arcmin
around the Galactic centre. The monitoring survey towards the Magellanic Clouds (3 deg$^2$ for the LMC and 1 deg$^2$ for the SMC) was done
over 13 years (Ita et al. 2009). Over 20,000 variables down to $J\sim 18.5$~mag, $H\sim 17$~mag, and $K_s\sim 16$~mag, were detected in the SMC, and a catalogue is being prepared for publication (Ita et al. in prep). The catalogue of LMC variables will also be compiled in the near future. The light-curves that will be published will be incorporated into our database of templates after the quality of each light-curve is verified. As an example, in Fig. \ref{fig:matsunaga} we present the $K_S$-band light-curves of two Mira (top panels) and two Cepheid (bottom panels) variables obtained with SIRIUS/IRSF.

\subsubsection{WFCAM}\label{sect:wfcam}

\begin{figure}
   \centering
   \includegraphics[width=0.45\textwidth]{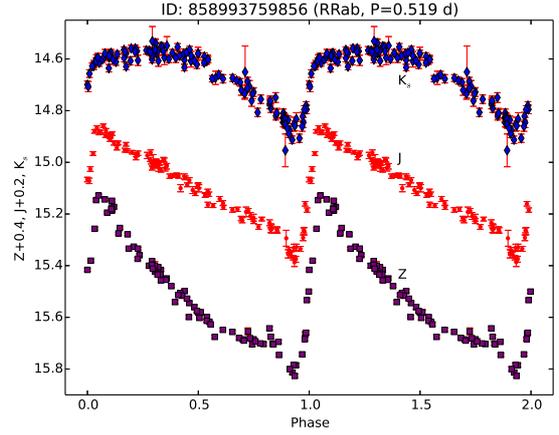}
      \caption{$Z,J,K$ phase-folded light-curves of object 858993759856 in the WSA. It is an RRab Lyrae star with a period of 0.51961138 days (from Ferreira Lopes et al., 2014). \label{fig:wfcam}}
   \end{figure}

The WFCAM Science Archive (WSA; Hambly et al. 2008; Cross et al. 2007, 2011) holds a large amount of unexploited public near-IR time-series data in its Calibration Database, that appears suitable for our purposes.
This archive contains five-band ($ZYJHK$) photometry for a total area of $\sim$10.4 square degrees of the sky made up of a few dozen scattered pointings, reaching $K\sim18$ mag. The detailed discussion of our comprehensive stellar variability analysis of the WFCAM Calibration Database, and the description of the resulting WFCAM Variable Star Catalog, will be given by Ferreira Lopes et al. (2014). We found a total of 49 variable stars already listed in the VSX (Watson et al. 2014) catalogue, most of them with unambiguous classifications based on optical data, and well-sampled near-IR photometry. These objects, together with a few dozen newly identified ones with certain classifications (based on visual inspection of their light-curves and their derived periods) were incorporated into our database of template light-curves. Figure \ref{fig:wfcam} shows, as an example, $Z,J,K$-band light-curves of an RRab star found in the WSA.

\section{The network of IR observatories}\label{sect:network}
With the aim of increasing the number of both high-quality light-curves and variability classes that could be incorporated into our near-IR templates database, we also undertook our own observational program, whose strategy was to monitor hundreds of optically well-studied variable stars in the $JHK_S$ bands. The astronomical community as a whole has been very supportive, with the end result that we have secured time for this project using several IR facilities across the globe. In order to ensure a homogeneous observational strategy and to optimize the use of the time awarded, each telescope/instrument combination was used to build template light-curves for at most a very few specific variability classes; for example, robotic (and those operated in service-mode) telescopes were mainly used to monitor bright field variables with periods $\gtrsim$1 day (Sects. \ref{sect:rem} \& \ref{sect:smarts}); short-period variables were observed with small- and medium-sized telescopes 
operated in visitor-mode to ensure the requested homogeneous 
phase 
coverage (Sects. \ref{sect:saao}, \ref{sect:irsf} \& \ref{sect:tcs}); facilities with large field-of-view cameras have been selected to monitor variables in clusters, i.e. relatively small areas of the sky with high concentration of (on average dimmer) variable stars (Sects. \ref{sect:boao} and \ref{sect:vista}).

In the next subsections we describe the facilities that joined the VVV Templates Project and present some sample light-curves that we have obtained in the last three years of observing campaigns. Table \ref{table:networkobs} summarizes the technical specifications of the IR cameras used in the project.

\begin{table*}
\caption{Technical specifications of the IR cameras used in the VVV Templates Project.}
\centering
    \begin{tabular}{ccccccccc}
	\hline \hline
         Imager       & REMIR & Mk II$^a$ & ANDICAM & SIRIUS  &    CAIN-III  & KASINICS & VIRCAM\\ 
 \hline
        Telescope		  & REM 0.6m  	& SAAO 0.75m	& SMARTS 1.3m	& IRSF 1.4m & TCS 1.5m & BOAO 1.8m & VISTA 4m\\
        Filter(s) Used 		  & $JK'$       & $K$		& $K$		&  $JK_S$   & $JHK_S$  &  $JHK_S$  & $JK_S$\\
        FOV (arcminutes)   	  & 10$\times$10& $\sim$12	& 2.4$\times$2.4& 7.7$\times$7.7       & 4.25$\times$4.25& 3.3$\times$3.3&66$\times$90\\ 
        Scale (arcseconds/pix)   & 1.2      	& -		& 0.28		&0.45       & 1        & 0.39 	 & 0.34	\\ 
        Gain (e-/ADU)            & 5        	& -		& 7.2		& 5         & 8.5      & 2.56 	 & 4.19$^b$	\\ 
        R.O.N. (e-)              & 100         & -		& 20		& 30        & 70       & 38 	 &  - $^b$	\\ 
        \hline
    \end{tabular}
    \label{table:networkobs}
    \begin{flushleft}
     {$a$}: InSb photometer;\\
     {$b$}: \texttt{http://casu.ast.cam.ac.uk/surveys-projects/vista/technical/vista-gain}\\
    \end{flushleft}
\end{table*}

\subsection{REM 0.6m telescope}\label{sect:rem}
The Rapid Eye Mount (REM) is a 60 cm completely robotic telescope located at ESO's La Silla Observatory. It hosts two instruments: ROSS2, a simultaneous multichannel imaging camera delivering Sloan $g', r', i', z'$ bands onto four quadrants of the same 2k$\times$2k CCD detector; and REMIR, an IR imaging camera equipped with a $J, H, K'$ filter set. The two cameras can observe the same 10$\times$10 arcmin$^2$ field of view simultaneously thanks to a dichroic before the telescope focus. The Observatory is operated for INAF by the REM Team\footnote{\tt http://www.rem.inaf.it/}.

\begin{figure}
   \centering
   \includegraphics[width=0.45\textwidth]{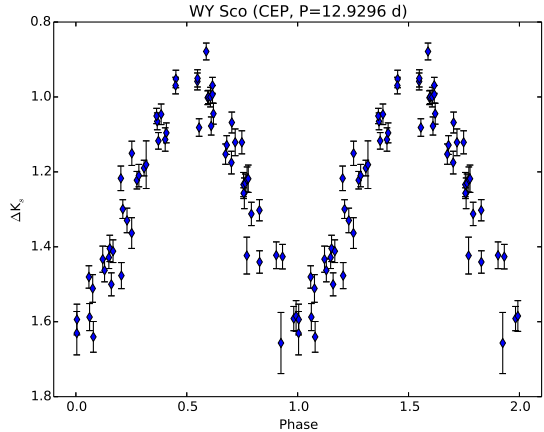}
      \caption{Differential $K'$-band light-curve of WY Sco, a W Virginis monitored with the REMIR camera at the 0.6m REM telescope in 2011.}
         \label{wysco}
   \end{figure}

Thanks to its fully robotic observing procedure and relatively large FOV, REM is a very suitable facility for multiwavelength monitoring of variable objects (it was conceived to provide prompt follow-up of GRBs, which it continues to deal with), and we have thus used it extensively for this project over the last semesters through observing time awarded by both INAF and Chilean Time Allocation Committees (TACs). Even though some technical problems did not always allow us to optimize the observing strategy, REM was capable of delivering template light-curves of a few medium-period ($\approx$tens of days) variable stars that would have otherwise been impossible to monitor via classical visitor-mode observing programs. As an example, in Fig. \ref{wysco} we present the $K'$-band light-curve of WY Sco, a W Virginis variable monitored with REMIR in 2011.

\subsection{SAAO 0.75m telescope}\label{sect:saao}
In September 2010 we observed for 21 nights with the 0.75m telescope at the South African Astronomical Observatory (SAAO).  The instrumentation available with this telescope includes Mk II\footnote{\tt http://old.saao.ac.za/facilities/instrumentation/\\infrared-photometer-mk-ii/}, an IR photometer designed primarily to cover the $JHKL$ bands. It uses an InSb cell cooled with nitrogen slush, and a sectored mirror focal-plane chopper. The acquisition program (called BRUCE) also provides online data reductions, making corrections for atmospheric extinction and for zero-point. Considering the observing strategy in relation to the photometer and the telescope diameter, we decided to mainly focus on very bright, short-period variable stars.

\begin{figure}
   \centering
   \includegraphics[width=0.45\textwidth]{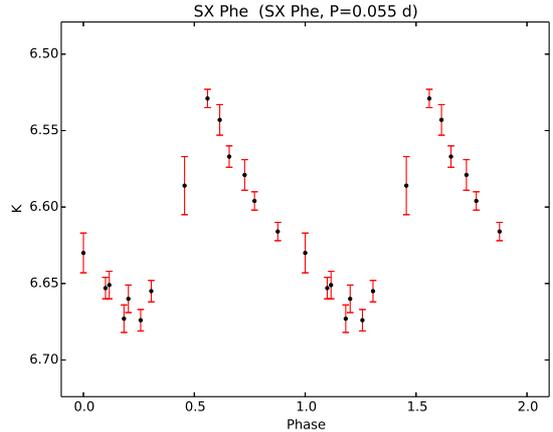}
      \caption{$K$-band light-curve of SX Phe, the prototype of the class, taken with the SAAO 0.75m telescope in September 2010.}
         \label{sxphe}
   \end{figure}

Unfortunately, only a few nights over the whole run were truly photometric, and this circumstance did not allow us to obtain complete phase coverage for several light-curves. Nonetheless, by virtue of the very small photometric error, the data delivered by Mk II produced a few excellent light-curves for this project. As an example, we show in Fig. \ref{sxphe} the light-curve of SX Phe, the prototype of the class. 

\subsection{SMARTS 1.3m telescope}\label{sect:smarts}
This 1.3m telescope, located in the premises of Cerro Tololo Inter-American Observatory (CTIO), Chile, was previously the 2MASS southern telescope before the SMARTS Consortium took over its operation\footnote{\tt http://www.ctio.noao.edu/noao/content/\\SMARTS-13-m-Telescope}, now entirely offered in service/queue mode. At the SMARTS 1.3m telescope a dual-channel, optical-IR imager (ANDICAM) is permanently mounted. This takes simultaneous $BVRI$ and $JHK$ data using a dichroic with a CCD and a Rockwell 1024$\times$1024 HgCdTe Hawaii Array.  The latter has a FOV of 2.4$\times$2.4 arcmin$^2$. A moveable mirror allows dithering in the IR while an optical exposure is going on.

\begin{figure}
   \centering
   \includegraphics[width=0.45\textwidth]{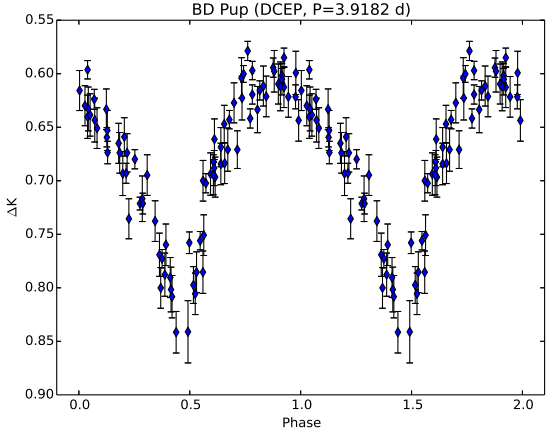}
      \caption{Differential $K$-band light-curve of BD Pup, a $\delta$ Cep variable star observed at the SMARTS 1.3m telescope during the 2011A semester.} 
         \label{bdpup}
   \end{figure}

Because of its service-mode observing procedure this facility is potentially very suitable for monitoring variable stars with periods longer than a few days. Unfortunately, however, the very small FOV of the IR array combined with the not always optimal pointing accuracy of the telescope have significantly decreased the value of this facility for our project; in particular, it was difficult in several cases to find a reliable and fixed set of reference stars over which to perform differential aperture photometry. Nonetheless, a few light-curves obtained at the SMARTS 1.3m telescope can still be considered of template-like quality. As an example, in Fig. \ref{bdpup} we present the $K$-band light-curve of BD Pup, a $\delta$ Cep variable star observed during the 2011A campaign.

\subsection{IRSF 1.4m telescope}\label{sect:irsf} 
\begin{figure}
   \centering
   \includegraphics[width=0.45\textwidth]{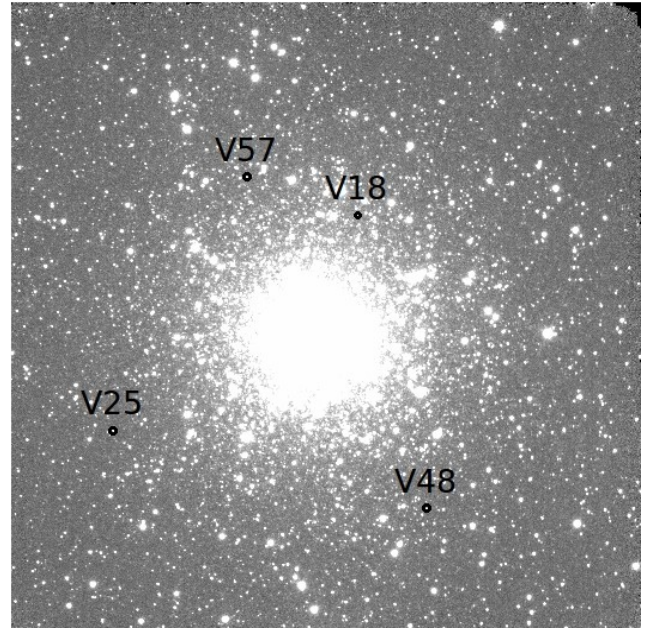}
      \caption{SIRIUS $K_S$-frame image of M62, observed at IRSF in Sept 2010. The FOV is 7$\times$7 arcminutes$^2$, north is up and east is to the right. Marked with black circles are the four RRab variables shown in Fig. \ref{fig:m62rrab}.}
         \label{fig:m62sirius}
   \end{figure}

\begin{figure}
   \centering
   \includegraphics[width=0.49\textwidth]{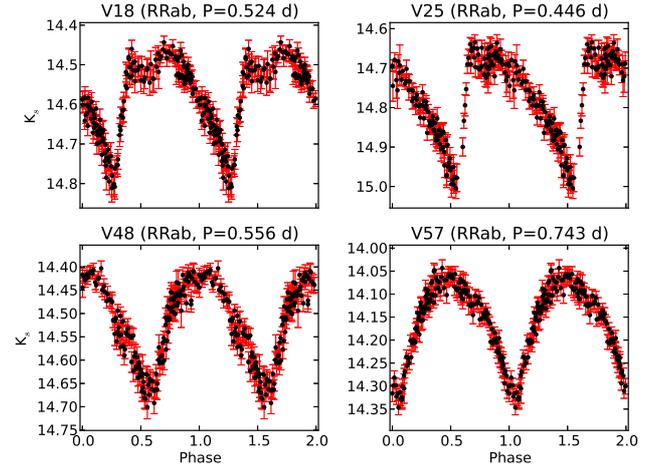}
      \caption{SIRIUS $K_S$-light-curves of four RRab variable stars in the globular cluster M62.}
         \label{fig:m62rrab}
   \end{figure}

The VVV Templates Project was able to observe with SIRIUS (see also Sect. \ref{sect:matsunaga}) during three-night-long runs in July and September 2010, and in March 2011. By virtue of the camera's relatively large FOV (especially when compared to other imagers at similar-sized telescopes; see Table \ref{table:obs}), we optimized the target selection by focusing on the globular cluster M62 (NGC~6266; see Fig. \ref{fig:m62sirius}), one of the most RR Lyrae-rich clusters in the Galaxy. It is known to contain more than 240 variables, of which 133 are fundamental-mode RR Lyrae stars (RRab), 76 first overtone (RRc) pulsators, 4 Type II Cepheids, and 25 long-period variables (Contreras et al. 2010). At the end of the M62 SIRIUS campaign, we collected over 150 data points in $JHK_S$. In Fig. \ref{fig:m62rrab} we show template-quality light-curves of four M62 RRab variable stars obtained with SIRIUS. 

We also observed the open cluster NGC~6134, which hosts six $\delta$ Sct stars; unfortunately, the overall amplitudes of these pulsators were smaller than 0.1 mag in $K_S$. Thus, we were 
unable to recover any suitable light-curves, this amplitude being of the same 
order of magnitude as the scatter of the comparison stars in the differential photometry.

\subsection{\textit{Carlos S\'{a}nchez} 1.5m telescope}\label{sect:tcs}
During the semesters 2010A-2011B we made intensive use of the Carlos S\'{a}nchez 1.5m Telescope (TCS), located in the IAC's (Instituto de Astrof\'isica de Canarias) Teide Observatory; TCS is equipped with CAIN-III, an IR imager built by the IAC\footnote{\tt http://www.iac.es/telescopes/pages/en/home/\\instruments/cain.php?lang=EN}. Its detector is composed of a mosaic of 256 x 256 HgCdTe photovoltaic elements (NICMOS 3 technology), sensitive in the 1.0-2.5 $\mu$m spectral range, together with four bias electronic devices, each of them controlling a quadrant and with an independent readout. Each pixel's physical size is 40 $\mu$m, corresponding to scales in the focal plane of 1 arcsec/pixel and 0.39 arcsec/pixel for the ``wide'' and ``narrow'' field configurations, respectively. Therefore, the FOV for these two configurations is 4.2$\times$4.2 arcminutes$^2$ (wide) and 1.7$\times$1.7 arcminutes$^2$ (narrow).

We observed for several weeks with the goal of recovering template-quality light-curves of short- to medium-period variable stars, mainly eclipsing binaries and $\delta$ Sct stars, the latter selected because they are known to be either (optically) high-amplitude $\delta$ Sct stars or members of relatively compact open clusters. In order to optimize the number of comparison stars in the FOV, we were forced to opt for the CAIN-III ``wide-field'' optical configuration. The ability to image an area of sky as big as 4.2 square arcminutes had the drawback of a 1 arcsec/pixel scale. Under good-seeing conditions, this relatively poor scale has in some cases prevented a proper sampling of the stellar PSF, and to some extent has compromised our ability to perform high-precision (i.e. of the order of a few mmag) aperture photometry. The data reduction was performed with \textit{caindr}, an IRAF package developed at IAC to specifically take into account the CAIN-III special features and image format.

\begin{figure}
   \centering
   \includegraphics[width=0.45\textwidth]{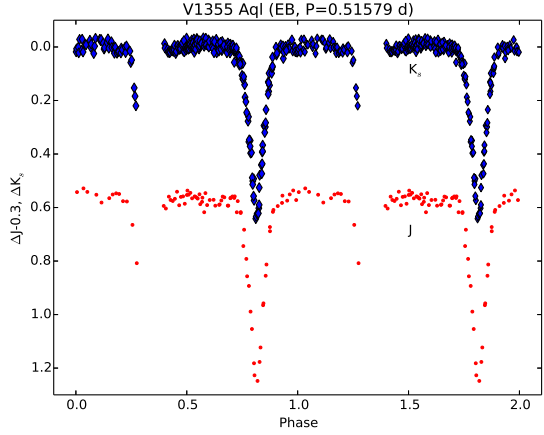}
      \caption{Differential $J$- and $K_S$-band light-curve of the Algol-type eclipsing binary V1355 Aql, obtained at the \textit{Carlos S\'{a}nchez} 1.5m Telescope in July 2011.
       Unfortunately the secondary eclipse was poorly covered because of bad weather.\label{v1355aqlk}}
   \end{figure}

In Fig. \ref{v1355aqlk} we present the resultant light-curve from the intensive monitoring in the $J$ (86 data points) and $K_S$ (252 data points) filters of V1355 Aql, an Algol-type eclipsing binary with an orbital period of 0.5158 day.

We note that in some cases we simultaneously observed the same target stars in the near-IR at TCS and in the optical at the 67/92 cm Schmidt telescope at the INAF Astronomical Observatory of Padova, Asiago (Italy), thus collecting multiband light-curves from the B to the K bands.

\subsection{BOAO 1.8m telescope}\label{sect:boao}
Every semester from late 2010 until the beginning of 2012, the VVV Templates Project was awarded 15 nights of visitor-mode observing time at the Bohyunsan Optical Astronomy Observatory (BOAO), in the Republic of Korea. The near-IR imager installed at the BOAO 1.8m telescope is named KASINICS (Moon et al. 2008)  and represents one of the KASI research projects to develop astronomical IR observation systems. It uses a 512$\times$512 InSb array (Aladdin III Quadrant, Raytheon Co.) to enable $J, H, K_S$, and $L$ band observations. The FOV of the array is 3.3$\times$3.3 arcminutes$^2$, with a resolution of 0.39 arcsec/pixel.

\begin{figure*}
   \centering
   \includegraphics[width=0.45\textwidth]{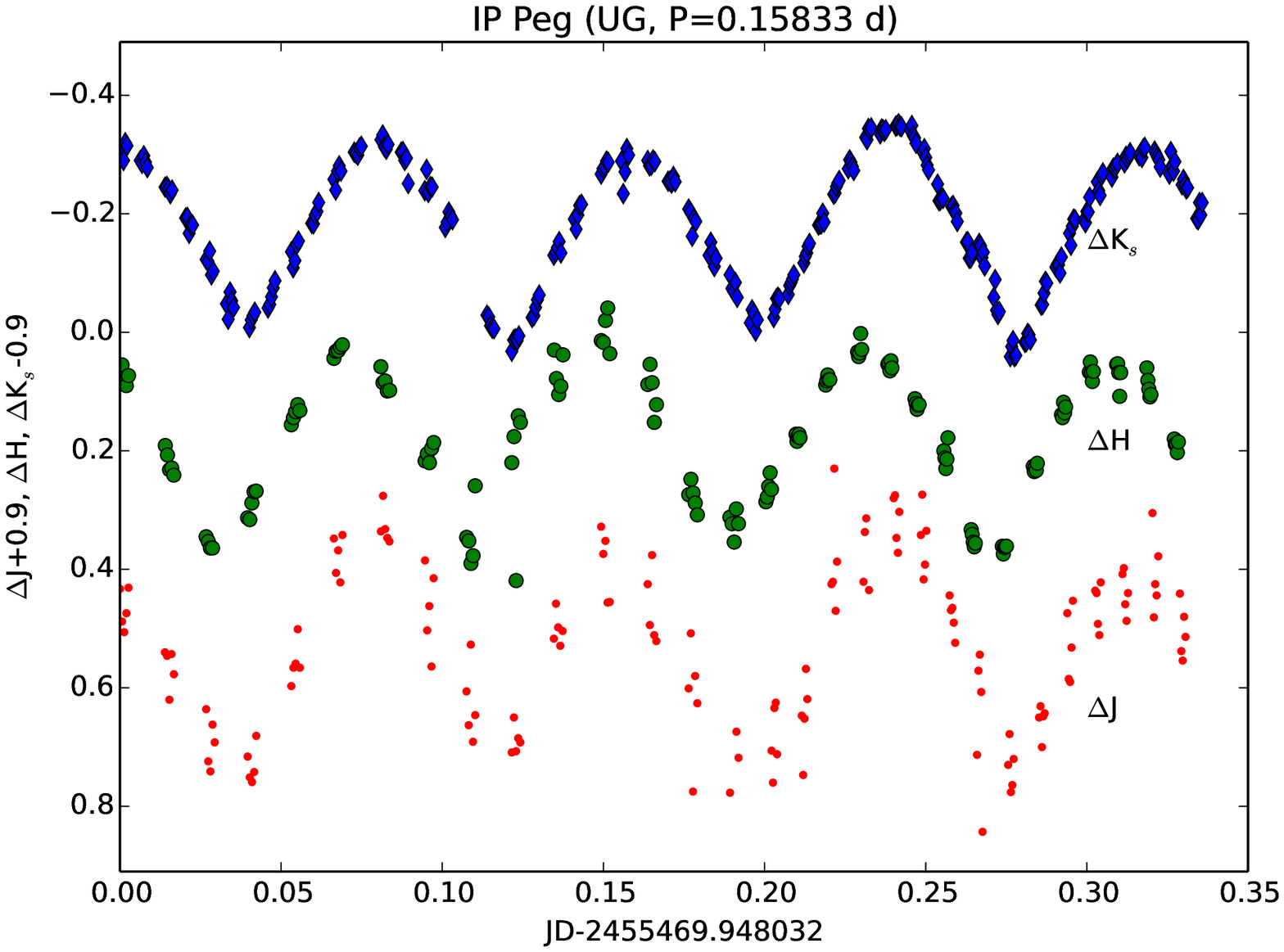}
   \includegraphics[width=0.45\textwidth]{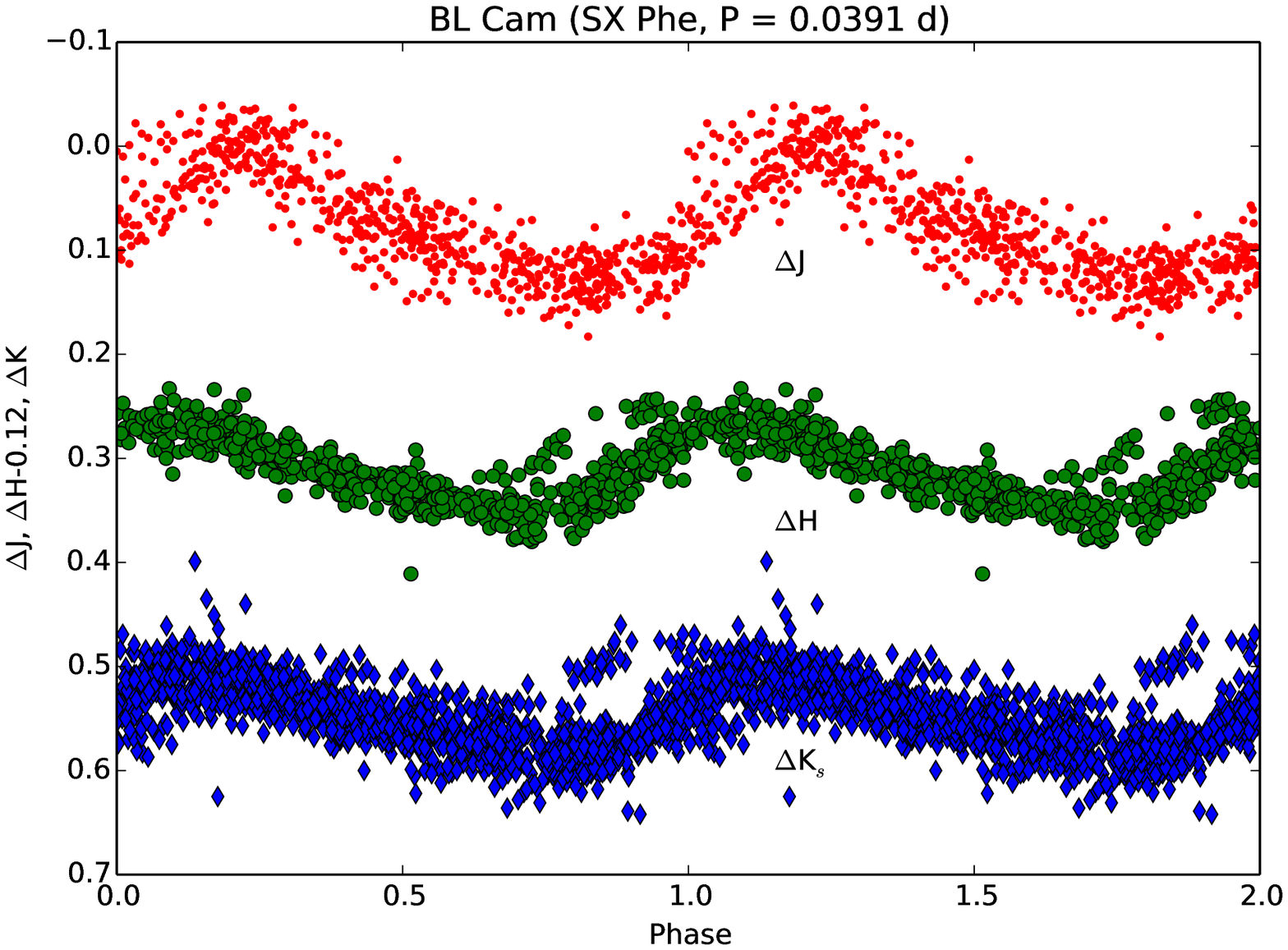}
   \caption{BOAO-KASINICS $JHK_S$ light-curves of IP Peg, a well-studied cataclysmic variable (left panel) and BL Cam, an extreme metal-deficient, high-amplitude SX Phe variable star (right panel).
         \label{boao}}
   \end{figure*}

The VVV Template targets selected for intensive photometric monitoring with this facility were mainly cataclysmic variables and moderately dim SX Phe and $\delta$ Sct stars. Although a significant fraction of the awarded time was unusable because of unfavourable weather conditions, especially during the rainy season, the data obtained at the BOAO 1.8m telescope led to some template light-curves with extremely uniform phase coverage, a large number of data points (1270 for the BL Cam $K_S$ light-curve; see Fig. \ref{boao}), and very small photometric errors. In Fig. \ref{boao} we show $JHK_S$ light-curves of two well-studied variable stars: the Cataclysmic Variable IP Peg (e.g. Ribeiro et al. 2007) and the extreme metal-deficient, high-amplitude SX Phe variable BL Cam (e.g. Fauvaud et al. 2010).

\subsection{VISTA 4m telescope}\label{sect:vista}
Starting with Period 87 ESO offered up to 25\% of the observing time with the VISTA 4m telescope to open PI projects (previously it had been exclusively dedicated to the ESO Public Surveys\footnote{\tt http://www.eso.org/public/teles-instr/\\surveytelescopes/vista/surveys/}). In P87, we were awarded 34 hours of service-mode to carry out VIRCAM time-series observations of $\omega$~Centauri (NGC~5139). At the end of the run (April 2012), we totalled 42 epochs in $J$ and 100 epochs in $K_S$, over a temporal baseline of 352 days.

The source $\omega$~Cen was identified as an excellent template target to be monitored with VISTA not only because of its rich variable star content (Clement et al. 2001) and large apparent size on the sky ($\sim$115 arcmin diameter; Harris 1996) that matches the huge VIRCAM FOV quite well, but also for its well-known intrinsic scientific interest.
This source is known to host more than 400 variable stars, among which are at least 195 RR Lyrae stars, $\sim$70 SX Phoenicis (the largest number found in any stellar system), more than 80 eclipsing binaries, at least 25 Long-Period variables, 7 Type II Cepheids (i.e. BL Her, W Vir, and RV Tau), a few spotted and ellipsoidal stars, and one cataclysmic variable candidate (Clement's catalogue, 2013 version\footnote{\tt http://www.astro.utoronto.ca/$\sim$cclement/read.html}). 

\begin{figure*}
   \centering
   \includegraphics[width=0.9\textwidth]{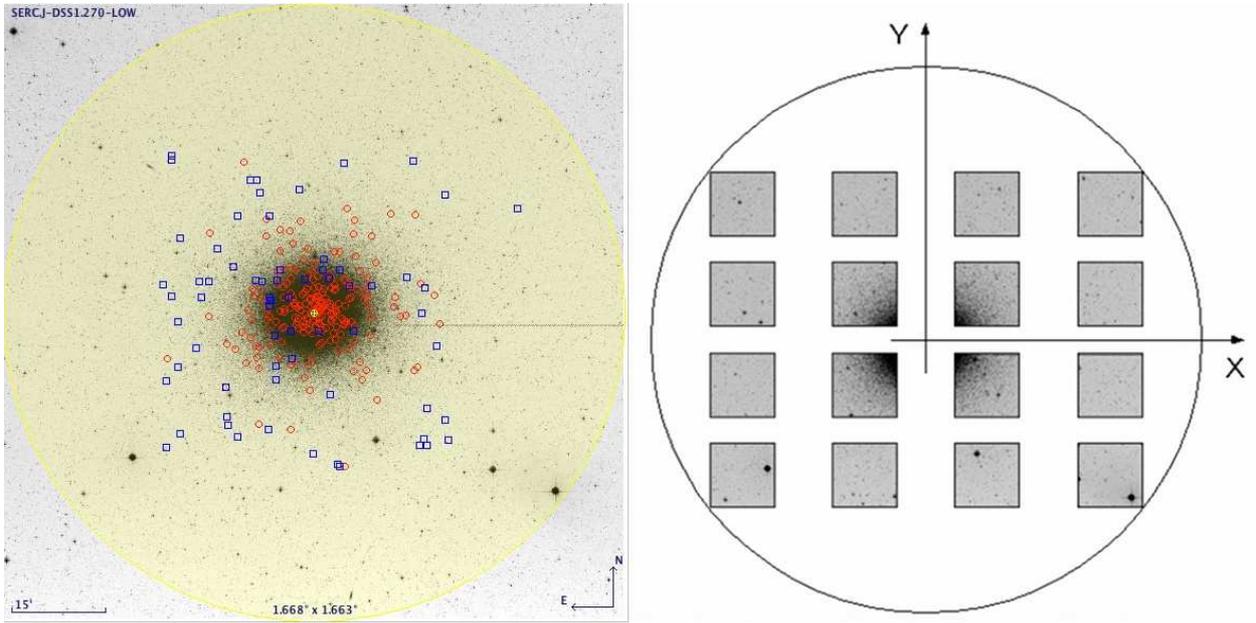}
      \caption{VISTA observations of $\omega$~Cen. Left panel: an example of variable star distribution across the sky field, where red circles mark the positions of known RR Lyrae stars, blue squares of known eclipsing binaries, and the yellow circle represents the VISTA FOV. Right panel: schematic map of VIRCAM's 16 detectors, with $\omega$~Cen at the centre of the focal plane.}
         \label{omegacenvvv}
   \end{figure*}

\begin{figure}
 \centering
 \includegraphics[width=0.47\textwidth]{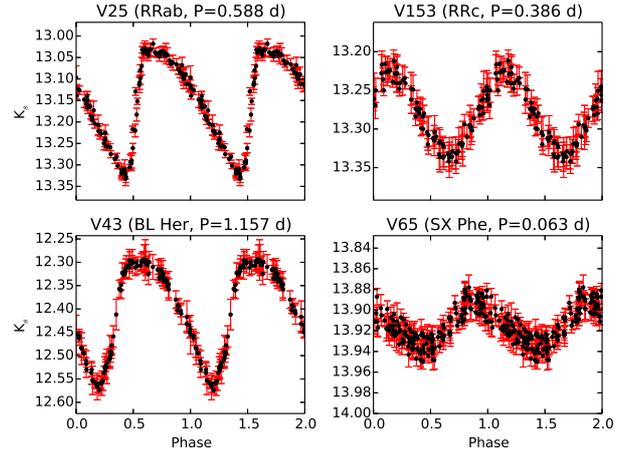}\\
 \caption{VISTA K$_S$-band light-curves of four stellar pulsators in $\omega$~Cen. Clockwise from the top left: a fundamental-mode RR Lyrae (RRab), a first overtone (RRc) RR Lyrae, an SX Phe, and a BL Her variable.}
 \label{omegapulsator}
\end{figure}

\begin{figure}
 \centering
 \includegraphics[width=0.47\textwidth]{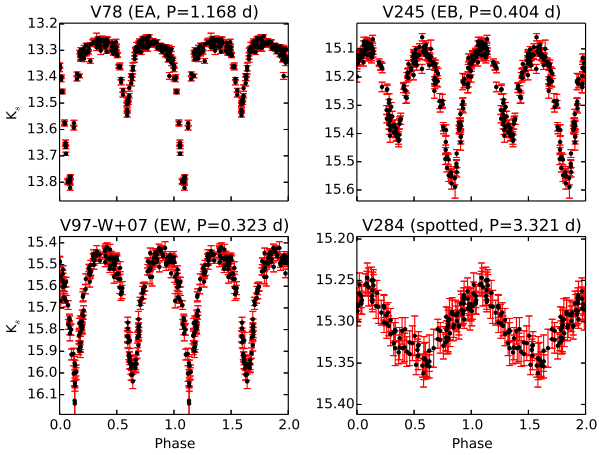}\\
 \caption{VISTA K$_S$-band light-curves of additional variable stars in $\omega$~Cen: V78 (top left), an Algol-type (EA) eclipsing binary; V245 (top right), a $\beta$ Lyrae-type (EB) eclipsing binary; V97-W07 (bottom left), a W Ursae Majoris-type (EW) eclipsing binary reported in Weldrake et al. (2007); V284 (bottom right), a spotted variable star.}
 \label{omegaeclips}
\end{figure}

With an effective field of view of 1.1 $\times$ 1.5 deg${}^{2}$ and a pixel scale of 0.34 arcsec/pix, VIRCAM has allowed us to monitor practically all the known variable stars (Fig. \ref{omegacenvvv}). Indeed, only five previously known variables associated with $\omega$~Cen in the current version of the Clement catalogue are outside the VISTA FOV: 4 RRab (V171, V178, V179, and V182) and V175, an unclassified variable.
Light-curves of the known variable stars were recovered using as references the catalogues of Clement et al. (2001), Kaluzny et al. (2004), Samus et al. (2009b), and Weldrake et al. (2007). All the coordinates and periods were revised; some inconsistencies and errors were found, mainly in the coordinates listed in the literature (see, e.g., Navarrete et al. 2013). The complete variable star catalogue with precise coordinates, periods, and $J$ and $K_{\rm s}$ magnitudes will be published in Navarrete et al. (in prep.). Some examples of template light-curves derived in this way are presented in Figs. \ref{omegapulsator} and \ref{omegaeclips}.

\section{VVV itself as a source of template light-curves}  \label{sect:vvv}
The sky surveyed by VVV has partial overlaps with a few optical surveys that in recent years have scanned the most crowded regions of the Milky Way searching mainly for microlensing events (e.g. MACHO and OGLE Projects). This means that some variable stars that have already been classified in the optical could, in principle, be included in our IR template archive when VVV has collected enough epochs for the corresponding light-curves to be considered of template-like quality. Catelan et al. (2013) present a review of the present status of the variability phase of VVV, including the number of epochs per tile, variations of epoch cadence, etc.\\

\begin{figure*}
   \centering
   \includegraphics[width=0.9\textwidth]{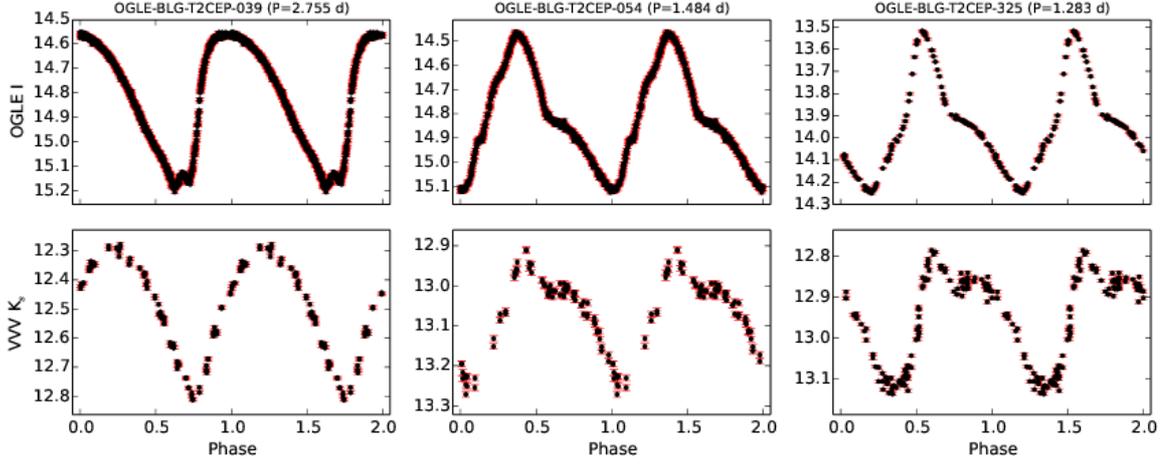}
      \caption{OGLE $I$-band (top panels) and VVV $K_S$-band light-curves (bottom panels) of three BL Her variables in the field. We note the change in amplitude and shape of the light-curve from the optical to the near-IR.}
         \label{gergely}
   \end{figure*}

A large fraction of the field sample comprises RR Lyrae, Cepheids (both Classical and Type II) and LPVs. As examples of optically well-characterized variable stars with VVV light-curves still under-construction (but already of very good quality), we show in Fig. \ref{gergely} three BL Her variables. These light-curves also illustrate that it is impossible to simply scale $I$-band light-curves as alternatives for $JHK$-band templates. 

\begin{figure}
  \includegraphics[width=0.45\textwidth]{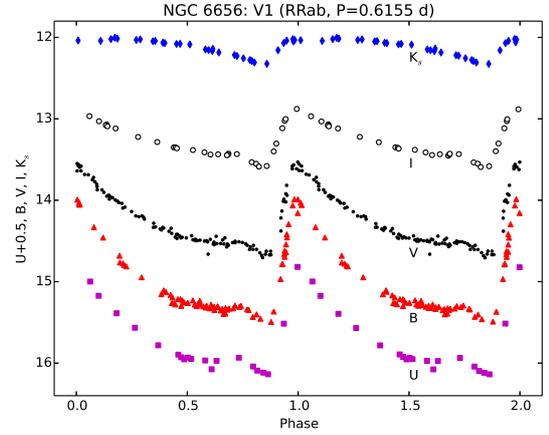}
  \caption{Light-curves at different wavelengths for NGC~6656 V1, a Known RRab-type star. 
    The $K_S$-band light-curve comes from our VVV data, while the $U$, $B$, $V$, and
    $I$ light-curves are from  Kunder et al. (2013). We also note the
    change in amplitude and shape of the light-curve at different
    wavelengths.}
  \label{javi1}
\end{figure}

Thirty-six known Galactic globular clusters (GCs from the Harris (1996) catalogue) also fall in the sky fields monitored by VVV, along with a few new candidates (Minniti et al. 2011, Moni Bidin et al. 2011). High extinction towards the low-latitude fields complicate their analysis, and the study of many of these clusters has been historically neglected, but there are several VVV GCs with a significant number of variables already detected (e.g. NGC~6441, NGC~6656, NGC~6638; e.g., Pritzl et al. 2001, Kunder et al. 2013). Although the number of epochs in VVV is generally not yet sufficient to provide us with template-quality light-curves for field variables, we do have enough to clearly see variability in numerous stars (Fig. \ref{javi1}). We also started a systematic search of variables in all the VVV GCs and have already found some new candidates in highly-reddened GCs including 2MASS-GC02 and Terzan 10 (Alonso-Garc\'{\i}a et al. 2013). For the latter, in particular, the number of epochs is almost 100 (Fig. \
ref{javi2}), and thus high enough to show the kind of data quality that we expect for all the VVV light-curves at the end of the survey. 
 

 \begin{figure}
\includegraphics[width=0.45\textwidth]{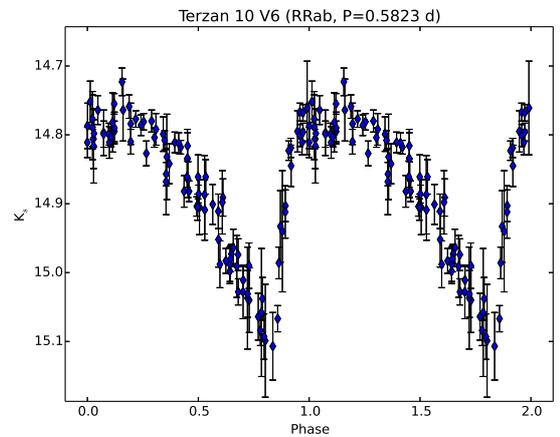}
\caption{Light-curve of one of the RR~Lyrae candidates found
  in Terzan~10.}
\label{javi2}
\end{figure}

\section{Concluding remarks}\label{sec:cr}
In the present study, we have introduced the VVV Templates Project, namely the coordinated effort towards the automated classification of the millions of light-curves that will be obtained over the course of the next few years from the VVV ESO Public Survey. Because VVV is the first massive, multi-epoch survey of stellar variability in the near-IR, the template light-curves that are needed in this band for training the automated variable star classification algorithms are not yet  available. Therefore, in this first paper in the VVV Templates series, we have described the construction of the first comprehensive database of stellar variability in the near-IR. 

We began by describing our efforts to collect data from the literature, and to mine public databases (such as 2MASS and WFCAM calibration catalogues) for high-quality near-IR light-curves. In addition, we described targeted observations at several IR facilities around the world of stars belonging to the various different variability classes.
The VVV database itself has now started to provide increasing numbers of high-quality light-curves of previously optically-classified variable stars, which makes our near-IR templates database a constantly growing dataset unique on its own. In Table \ref{tab:archive} we show the approximate numbers of template light-curves per variability class that  we have been able to gather in the VVV Templates Project, at the time of writing. 

Clearly, the scientific return of the VVV Templates Project is not restricted to the automated classification of the light-curves
provided by the VVV Survey. Rather, this dataset is providing us with a unique opportunity to expand our knowledge of the
stellar variability phenomenon per se, by tackling the comparatively ill-explored near-IR regime. 

The development, testing, and optimization of the automated classification algorithms of VVV light-curves will be the topic of the forthcoming papers of the series (see Catelan et al. 2013 for a preview), along with the scientific results arising from the systematic analysis of the most valuable datasets that we have been able to collect.

Our database can be publicly accessed through our dedicated web page (see footnote 1). Most of the data are already available and the rest will be released within at most two years from the publication of this paper.

\begin{table}
\caption{Approximate numbers of template light-curves (NLC) per variability class in the VVV Templates Project database, as of the time of writing.\label{tab:archive}}
\centering
    \begin{tabular}{ccccccccc}
	\hline \hline
 Variability class   & NLC \\
 \hline
        Cataclysmic variables (CV)  & 5 \\
        Classical Cepheids (DCEP) & 260 \\
        Type-II Cepheids (CW) & 230 \\
        Anomalous Cepheids (BLBOO) & 48  \\
        Delta Scuti (DSCT) & 2 \\
        Eclipsing Binaries (EA+EB+EW) & 100 \\ 
        Long-Period Variables (LPV)  & 1000 \\ 
        Orion Variables (IN) & 1\\
        Quasi-Stellar Objects (QSO) & 70\\
        R Coronae Borealis (RCB)   & 12 \\
        RR Lyrae, type RRab (RRab)  & 1600 \\
        RR Lyrae, type RRc (RRc) & 90 \\
        SX Phoenicis (SXPHE) & 9\\
        T Tauri stars (INT)  & 1 \\
        Wolf-Rayet stars (WR) & 6 \\
        Young Stellar Objects (YSO)  & 80\\
        X-ray Binaries (X) & 2 \\
        \hline
    \end{tabular}
    \label{table:obs}
\end{table}

\begin{acknowledgements}
This project is supported by the Ministry for the Economy, Development, and Tourism's Programa Iniciativa Cient\'ifica Milenio through grant
IC120009, awarded to the Millennium Institute of Astrophysics (MAS), by Proyecto Basal PFB-06/2007, and also by the IAC and by the Ministry of Science and Competitiveness of the Kingdom of Spain (grant AYA2010-16717). M.C., I.D., F.G., C.N., and J.A.G. acknowledge support by Proyectos Fondecyt Regulares \#1110326, \#1130196 and \#1141141.
M.C., S.E., A.J. and K.P. acknowledge support from the Vicerrector\'ia de Investigaci\'on (VRI), Pontificia Universidad Cat\'olica de Chile (Proyecto Investigaci\'on Interdisciplinaria 25/2011).
R.C.R. also acknowledges support by Proyecto Fondecyt Postdoctoral \#3130320.
F.G. also acknowledges support from CONICYT-PCHA Mag\'ister Nacional \#2014-22141509.
J.A.G. also acknowledges support by Proyecto Fondecyt Postdoctoral \#3130552.
C.N. also acknowledges financial support from CONYCIT-PCHA/Mag\'ister Nacional \#2012-22121934. 
J.C.B. acknowledges support by CONICYT-PCHA/Doctorado Nacional\#2013-21130248.
JB received support from FONDECYT Regular \#1120601.
RdG acknowledges research support from the National Natural Science Foundation of China (NSFC) through grant \#11373010.
C.E.F.L. acknowledges support from CNPq/Brazil through projects 150495/2014-5.
K.G.H. acknowledges support provided by the National Astronomical Observatory of Japan as Subaru Astronomical Research Fellow, and by the Polish National Science Center through grant 2011/03/N/ST9/01819. We kindly acknowledge the support provided by Schmidt Telescope (Asiago, Italy) of INAF - Osservatorio Astronomico di Padova (Large program 2012-2013 - P.I. Angeloni).
R.K.S. acknowledges support from CNPq/Brazil through projects 310636/2013-2 and 481468/2013-7. 
PAW thanks the SA~NRF for a research grant.
This research has made use of the International Variable Star Index (VSX) database, operated at AAVSO, Cambridge, Massachusetts, USA. This research has also made use of the SIMBAD database, operated at CDS, Strasbourg, France.
\end{acknowledgements}

\end{document}